\documentclass[11pt,fleqn]{article}
\usepackage{amsfonts, epsfig, amssymb, amsmath}
\usepackage[normalem]{ulem}
\usepackage{color}
\usepackage{tikz}
\usepackage{pgfplots}
\usepackage{float}

\newcommand{\ol}{\setlength{\itemsep}{0pt.}\begin{enumerate}}
\newcommand{\eol}{\end{enumerate}\setlength{\itemsep}{-\parsep}}
\newcommand{\ignore}[1]{}
\title{On codes decoding a constant fraction of errors on the BSC}
\author{Alex Samorodnitsky\thanks{School of Engineering and Computer Science,
The Hebrew University of Jerusalem,
Jerusalem 91904, Israel. Research partially supported by ISF
grant 1724/15.}~ and Ori Sberlo\thanks{Department of Computer Science, Tel Aviv University, Tel Aviv, Israel. The research leading to these results has received funding from the Israel Science Foundation (grant number 552/16) and from the Len Blavatnik and the Blavatnik Family foundation.
}}

\begin{document}
\date{}
\maketitle


\newtheorem{THEOREM}{Theorem}[section]
\newenvironment{theorem}{\begin{THEOREM} \hspace{-.85em} {\bf :}
}%
                        {\end{THEOREM}}
\newtheorem{LEMMA}[THEOREM]{Lemma}
\newenvironment{lemma}{\begin{LEMMA} \hspace{-.85em} {\bf :} }%
                      {\end{LEMMA}}
\newtheorem{COROLLARY}[THEOREM]{Corollary}
\newenvironment{corollary}{\begin{COROLLARY} \hspace{-.85em} {\bf
:} }%
                          {\end{COROLLARY}}
\newtheorem{PROPOSITION}[THEOREM]{Proposition}
\newenvironment{proposition}{\begin{PROPOSITION} \hspace{-.85em}
{\bf :} }%
                            {\end{PROPOSITION}}
\newtheorem{DEFINITION}[THEOREM]{Definition}
\newenvironment{definition}{\begin{DEFINITION} \hspace{-.85em} {\bf
:} \rm}%
                            {\end{DEFINITION}}
\newtheorem{EXAMPLE}[THEOREM]{Example}
\newenvironment{example}{\begin{EXAMPLE} \hspace{-.85em} {\bf :}
\rm}%
                            {\end{EXAMPLE}}
\newtheorem{CONJECTURE}[THEOREM]{Conjecture}
\newenvironment{conjecture}{\begin{CONJECTURE} \hspace{-.85em}
{\bf :} \rm}%
                            {\end{CONJECTURE}}
\newtheorem{MAINCONJECTURE}[THEOREM]{Main Conjecture}
\newenvironment{mainconjecture}{\begin{MAINCONJECTURE} \hspace{-.85em}
{\bf :} \rm}%
                            {\end{MAINCONJECTURE}}
\newtheorem{PROBLEM}[THEOREM]{Problem}
\newenvironment{problem}{\begin{PROBLEM} \hspace{-.85em} {\bf :}
\rm}%
                            {\end{PROBLEM}}
\newtheorem{QUESTION}[THEOREM]{Question}
\newenvironment{question}{\begin{QUESTION} \hspace{-.85em} {\bf :}
\rm}%
                            {\end{QUESTION}}
\newtheorem{REMARK}[THEOREM]{Remark}
\newenvironment{remark}{\begin{REMARK} \hspace{-.85em} {\bf :}
\rm}%
                            {\end{REMARK}}

\newcommand{\thm}{\begin{theorem}}
\newcommand{\lem}{\begin{lemma}}
\newcommand{\pro}{\begin{proposition}}
\newcommand{\dfn}{\begin{definition}}
\newcommand{\rem}{\begin{remark}}
\newcommand{\xam}{\begin{example}}
\newcommand{\cnj}{\begin{conjecture}}
\newcommand{\mcnj}{\begin{mainconjecture}}
\newcommand{\prb}{\begin{problem}}
\newcommand{\que}{\begin{question}}
\newcommand{\cor}{\begin{corollary}}
\newcommand{\prf}{\noindent{\bf Proof:} }
\newcommand{\ethm}{\end{theorem}}
\newcommand{\elem}{\end{lemma}}
\newcommand{\epro}{\end{proposition}}
\newcommand{\edfn}{\bbox\end{definition}}
\newcommand{\erem}{\bbox\end{remark}}
\newcommand{\exam}{\bbox\end{example}}
\newcommand{\ecnj}{\bbox\end{conjecture}}
\newcommand{\emcnj}{\bbox\end{mainconjecture}}
\newcommand{\eprb}{\bbox\end{problem}}
\newcommand{\eque}{\bbox\end{question}}
\newcommand{\ecor}{\end{corollary}}
\newcommand{\eprf}{\bbox}
\newcommand{\beqn}{\begin{equation}}
\newcommand{\eeqn}{\end{equation}}
\newcommand{\wbox}{\mbox{$\sqcap$\llap{$\sqcup$}}}
\newcommand{\bbox}{\vrule height7pt width4pt depth1pt}
\newcommand{\qed}{\bbox}
\def\sup{^}

\def\H{\{0,1\}^n}

\def\S{S(n,w)}

\def\g{g_{\ast}}
\def\xop{x_{\ast}}
\def\y{y_{\ast}}
\def\z{z_{\ast}}

\def\f{\tilde f}

\def\n{\lfloor \frac n2 \rfloor}

\def \E{\mathop{{}\mathbb E}}
\def \R{\mathbb R}
\def \Z{\mathbb Z}
\def \F{\mathbb F}
\def \T{\mathbb T}

\def \x{\textcolor{red}{x}}
\def \r{\textcolor{red}{r}}
\def \Rc{\textcolor{red}{R}}

\def \noi{{\noindent}}

\def \iff{~~~~\Leftrightarrow~~~~}

\def \queq {\quad = \quad}

\def\<{\left<}
\def\>{\right>}
\def \({\left(}
\def \){\right)}

\def \e{\epsilon}
\def \l{\lambda}

\def\Tp{Tchebyshef polynomial}
\def\Tps{TchebysDeto be the maximafine $A(n,d)$ l size of a code with distance $d$hef polynomials}
\newcommand{\rarrow}{\rightarrow}

\newcommand{\larrow}{\leftarrow}

\overfullrule=0pt
\def\setof#1{\lbrace #1 \rbrace}

\begin{abstract}

Using techniques and results from \cite{KKMPSU} we strengthen the bounds of \cite{S} on the weight distribution of linear codes achieving capacity on the BEC. In particular, we show that for any doubly transitive binary linear code $C \subseteq \H$ of rate $0 < R < 1$ with weight distribution $\(a_0,...,a_n\)$ holds $a_i \le 2^{o(n)} \cdot \(1-R\)^{-2 \ln 2 \cdot \min\{i, n-i\}}$.

For doubly transitive codes with minimal distance at least $\Omega\(n^c\)$, $0 < c \le 1$, the error factor of $2^{o(n)}$ in this bound can be removed at the cost of replacing $1-R$ with a smaller constant $a = a(R,c) < 1- R$. Moreover, in the special case of Reed-Muller codes, due to the additional symmetries of these codes, this error factor can be removed at essentially no cost.

This implies that for any doubly transitive code  $C$ of rate $R$ with minimal distance at least $\Omega\(n^c\)$, there exists a positive constant $p = p(R,c)$ such that $C$ decodes errors on $\mathrm{BSC}(p)$ with high probability if $p < p(R,c)$. For doubly transitive codes of a sufficiently low rate (smaller than some absolute constant) the requirement on the minimal distance can be omitted, and hence this critical probability $p(R)$ depends only on $R$. Furthermore, $p(R) \rarrow \frac12$ as $R \rarrow 0$.

In particular, a Reed-Muller code $C$ of rate $R$ decodes errors on $\mathrm{BSC}(p)$ with high probability if
\[
R ~<~ 1 - \big(4p(1-p)\big)^{\frac{1}{4 \ln 2}},
\]
answering a question posed in \cite{AHN}.

\end{abstract}

\section{Introduction}

\noi The paper \cite{S} gave bounds on the weight distribution of linear codes achieving capacity on the binary erasure channel (BEC). In particular it was shown (\cite{S}, Proposition 1.6) that a  binary linear code $C$ of rate $R$ with weight distribution $\(a_0,...,a_n\)$ achieving capacity on the BEC under block-MAP decoding holds
\[
a_i ~\le~ 2^{o(n)} \cdot \(\frac{1}{1-R}\)^{2 \ln 2 \cdot \min\{i, n-i\}}.
\]

\noi The results of \cite{KKMPSU} imply that these bounds hold, in particular, for binary Reed-Muller codes.

\noi In this paper we strengthen the bounds above in two ways. We note that this improvement comes from taking a closer look at the results and the methods of \cite{KKMPSU}.

\noi First, we show the bounds in \cite{S} to hold for codes achieving capacity on the BEC under {\it bit-MAP decoding}. The results of \cite{KKMPSU} then imply that these bounds hold for any doubly transitive binary linear code.

\pro
\label{pro:BEC-bit-capacity}

\noi Let $C$ be a doubly transitive binary linear code of rate $R$. Let $\(a_0,...,a_n\)$ be the weight distribution of $C$. For $0 \le i \le n$, let $i^{\ast} = \min\{i,n-i\}$. Let $\theta = R^{2 \ln 2}$.

\begin{itemize}

\item For all $0 \le i \le n$ holds
\[
a_i ~\le~ 2^{o(n)} \cdot \(\frac{1}{1-R}\)^{2 \ln 2 \cdot i^{\ast}}.
\]

\item For all $0 \le i \le n$ holds
\[
a_i ~\le~ 2^{o(n)} \cdot \left\{\begin{array}{ccc} \frac{|C|}{(1-\theta)^{i^{\ast}} (1+\theta)^{n-i^{\ast}}} & 0 \le i^{\ast} \le \frac{1 - \theta}{2} \cdot n \\ \frac{{n \choose {i^\ast}} \cdot |C|}{2^n} & \mathrm{otherwise} \end{array} \right.
\]
\end{itemize}

\epro

\rem
\label{rem:binomial-dd}

\noi In particular, the second of these bounds implies that the weight distribution of a doubly transitive binary linear code of rate $R$ is essentially upper-bounded by that of a random code of the same rate in the band of weights of width $R^{2 \ln 2}$ around $\frac n2$. (Cf. \cite{Krasikov-Litsyn}, where similar behavior was inferred for codes with large dual distance.)

\erem

\noi Next, we observe that these bounds can be made more precise\footnote{Note that in the bounds below we replace $i^{\ast} = \min\{i,n-i\}$ with $i$. This slightly weakens the bounds (and can be avoided, at least for Reed-Muller codes) but does not affect the performance of a code on the BSC.} for codes whose minimal distance is somewhat large, depending on the rate of the code. We focus on the first bound, since it seems to be more relevant for the performance of a code on the BSC.

\pro
\label{pro:min-dist}

\noi We use the notation from Proposition~\ref{pro:BEC-bit-capacity}.

\begin{itemize}

\item Let $C$ be a binary Reed-Muller code of positive rate $0 < R < 1$. There exists $R^{\ast}$ such that $|R^{\ast} - R| \le o_n(1)$ and such that for all $0 \le i \le n$ holds

\[
a_i ~\le~ O\(\(1-R^{\ast}\)^{-2 \ln 2 \cdot i}\).
\]

\item Let $0 < R < 1$ and let $0 < c \le 1$ be constants. Let $C$ be a doubly transitive binary linear code of rate $R$ and minimal distance $\Omega\(n^c\)$. Then there exists a constant $a = a(R,c) < 1- R$, such that for all $0 \le i \le n$ holds
\[
a_i ~\le~ O\(a^{-2 \ln 2 \cdot i}\).
\]

\item Moreover, there exists an absolute constant $R_0 > 0$ so that if $R \le R_0$ and if $C$ is a doubly transitive binary linear code of rate $R$, there exists a constant $a = a(R) \ge 1 - R^{\Omega(1)}$, such that the inequality above holds for all $0 \le i \le n$.

\end{itemize}

\epro

\cor
\label{cor:RM_BSC}

\begin{itemize}

\item Let $C$ be a binary Reed-Muller code of positive rate $0 < R< 1$. Then $C$ decodes errors on $\mathrm{BSC}(p)$ with high probability (more precisely, a family of such codes $\{C_n\}_n$ with $\limsup_n R\(C_n\) \le R$, attains vanishing error probability on $\mathrm{BSC}(p)$ as $n \rarrow \infty$) if
\[
R ~<~ 1 - \big(4p(1-p)\big)^{\frac{1}{4 \ln 2}}.
\]

\item Let $0 < R < 1$ and let $0 < c \le 1$ be constants. Let $C$ be a doubly transitive binary linear code of rate $R$ and minimal distance $\Omega\(n^c\)$. Then $C$ decodes errors on $\mathrm{BSC}(p)$ with high probability if
\[
a ~>~ \big(4p(1-p)\big)^{\frac{1}{4 \ln 2}},
\]
where $a = a(R,c)$ is the constant from the second claim of Proposition~\ref{pro:min-dist}.

\item Let $R_0$ be the constant from the second claim of Proposition~\ref{pro:min-dist}. Let $R \le R_0$ and let $C$ be a doubly transitive binary linear code of rate $R$. Then $C$ decodes errors on $\mathrm{BSC}(p)$ with high probability if
\[
a ~>~ \big(4p(1-p)\big)^{\frac{1}{4 \ln 2}},
\]
where $a = a(R) \ge 1 - R^{\Omega(1)}$ is the constant from the third claim of Proposition~\ref{pro:min-dist}. In particular, $p \rarrow \frac12$ as $R \rarrow 0$.

\end{itemize}

\ecor

\noi The first claim of this corollary answers a question from \cite{AHN} (see also the discussion there). The third claim of the corollary says that any doubly transitive binary linear code of a sufficiently small rate performs well on the BSC.

\noi A well-known conjecture in information theory states that Reed-Muller codes achieve capacity on the BSC. This conjecture would be true if the RHS of the inequality in the first claim of this corollary would be replaced with $1-H(p)$, where $H(x) = -x\log_2(x) - (1-x) \log_2(1-x)$ is the binary entropy function. The next figure shows the two functions $1 - \big(4p(1-p)\big)^{\frac{1}{4 \ln 2}}$ and $1-H(p)$.


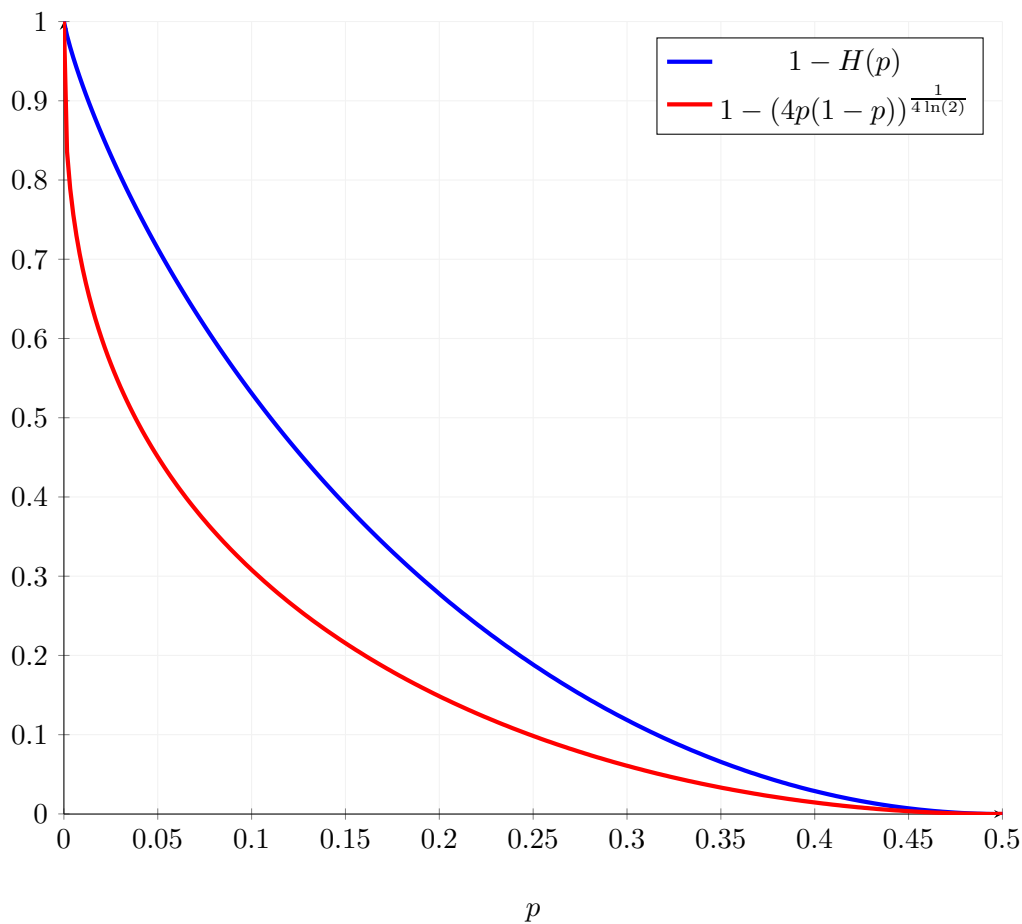
\begin{figure}[H]
	\centering
	\caption{The bound of Corollary~\ref{cor:RM_BSC} vs. the channel capacity}
	\label{fig: capacity vs our result}
	\begin{tikzpicture}
	\begin{axis}[
	axis lines = left,
	x label style={at={(axis description cs:0.5,-0.05)},anchor=north},
	y label style={at={(axis description cs:-0.025,0.5)},anchor=west},
	xlabel =  $\newline p$,
	grid=major, grid style={gray!10},
	width = 400pt,
	ticklabel style={
		/pgf/number format/fixed,
		/pgf/number format/precision=5
	},
	]
	
	\addplot[domain=0:0.5,
	samples=300,
	ultra thick,
	color=blue
	]
	{1+x*ln(x)/ln(2)+(1-x)*ln(1-x)/ln(2)};
	\addlegendentry{$1-H(p)$}
	
	\addplot[domain=0:0.5,
	samples=300,
	ultra thick,
	color=red
	]
	{1-(4*x*(1-x))^(1/ln(16))};
	\addlegendentry{$1-(4p(1-p))^{\frac{1}{4\ln(2)}}$}
	
	\end{axis}
	\end{tikzpicture}
\end{figure}

\section{Proofs}

\subsection{Proof of Proposition~\ref{pro:BEC-bit-capacity}}

\noi We refer freely to the notation in \cite{KKMPSU} and in \cite{S}. Let $C$ be a linear code of rate $R$. Let $r(\cdot) = r_C(\cdot)$ denote the rank function of the binary matroid defined by $C$. That is, $r_C(T)$ is the rank of the column submatrix of a generating matrix of $C$ which contains columns indexed by $T$. Let $f(S) = |S| - r(S)$ be a function on $\H$. For $0 \le \l \le 1$, let $\mu(\l) = \mu_C(\l) = \E_{S \sim \l} f(S)$.

\noi For $1 \le i \le n$, let $h_i(p)$ be the $i$'th EXIT function for $C$ on $\mathrm{BEC}(p)$ and let $h(p) = \frac 1n \sum_{i=1}^n h_i(p)$ be average exit function.

\noi The proof of the proposition is based on the following observation.

\lem
\label{lem:mu-p-der}
\[
\frac{d \mu}{d \l} ~=~ n \big(1 - h(1-\l)\big).
\]
\elem

\prf

\noi By the Margulis-Russo formula, we have that
\[
\frac{d \mu}{d\l} ~=~ \frac{1}{\l} \cdot \E_{S \sim \l} \sum_{i \in S} \big(f(S) - f(S \setminus i)\big) ~=~ \frac{1}{\l} \cdot \sum_{i=1}^n \sum_{S: i \in S} \l^{|S|} (1-\l)^{n-|S|} \big(f(S) - f(S \setminus i)\big).
\]

\noi Note that $f(S) - f(S \setminus i) = 1 - r(S) + r(S \setminus i) = \left\{\begin{array}{ccc} 1 & \mathrm{if} & r(S) = r(S \setminus i) \\ 0 & \mathrm{if} & r(S) > r(S \setminus i) \end{array} \right.$. Note also that given that $R = S \setminus i$ is the set of the coordinates which are not erased by the channel, we can recover the $i$-th bit iff $r(S) = r(S \setminus i)$. Hence, denoting by $P_{b,i}(p)$ the erasure probability for bit $i$ in $C$ on $\mathrm{BEC}(p)$ (as in \cite{KKMPSU}) with noise $p = 1-\l$, we have that
\[
P_{b,i}(p) ~=~ P_{b,i}(1-\l) ~=~\sum_{R: i \not \in R} \l^{|R|} (1-\l)^{n - |R|} 1_{r(R) < r(R \cup i)} ~=~
\]
\[
\sum_{R: i \not \in R} \l^{|R|} (1-\l)^{n - |R|} \Big(1 - f(R \cup i) - f(R)\Big) ~=~ (1-\l) - \frac{1-\l}{\l} \cdot \sum_{S: i \in S} \l^{|S|} (1-\l)^{n-|S|} \big(f(S) - f(S \setminus i)\big).
\]

\noi Recall that (see \cite{KKMPSU}) we have $p h_i(p) = P_{b,i}(p)$. Hence,
\[
\frac{d \mu}{d\l} ~=~ \sum_{i=1}^n \(1 - \frac{1}{1-\l} P_{b,i}(1-\l)\) ~=~ n - \sum_{i=1}^n h_i(1-\l) ~=~ n \cdot (1 - h(1-\l)).
\]
\eprf

\noi Theorem 12 in \cite{KKMPSU} states that a doubly transitive code $C$ achieves capacity on the BEC under {\it bit-MAP} decoding. This is observed to be equivalent to the fact that, assuming the rate of $C$ is $R$, the average EXIT function $h$ has a sharp threshold at $1-R$. This means that for $p > 1 - R + o_n(1)$ holds $h(p) = 1 - o_n(1)$ and for $p < 1 - R - o_n(1)$ holds $h(p) = o_n(1)$. By Lemma~\ref{lem:mu-p-der} this means that for such a code we have
$\mu'(\l) = 1 - o_n(1)$ for $\l > R + o_n(1)$, and $\mu'(\l) = o_n(1)$ for $\l < R - o_n(1)$. This, in particular, implies that
$\mu(R) ~=~ o(n)$.

\noi We can now conclude the proof of Proposition~\ref{pro:BEC-bit-capacity}. Since $\mu_C(\l) = \l n - \E_{S \sim \l} r_C(S)$, and since $\mu_C(R) = o(n)$, Proposition 1.3 and Lemma 1.4 in \cite{S} imply the second claim of the proposition precisely in the way they imply the claim of Proposition 1.6 in \cite{S}. Next, observe that if $C$ is doubly transitive, then so is $C^{\perp}$. Hence the argument above applies to $C^{\perp}$, and we have that $\mu_{C^{\perp}}(1-R) = o(n)$. We now proceed in the same way to derive the first claim of the proposition.

\eprf

\subsection{Proof of Proposition~\ref{pro:min-dist}}

\noi We start with the first claim of the proposition. Let $C$ be a Reed-Muller code of rate $0 < R < 1$. By \cite{Bourgain-Kalai}, see also the proof of Theorem 17 in \cite{KKMPSU}, we have that for the average EXIT function $h$ of a Reed-Muller code holds, for some absolute constant $c$ and for any $p$ bounded away from $0$ and $1$ that
\[
\frac{dh}{dp} ~\ge~ c \log(n) \log \log(n) \cdot h (1-h).
\]

\noi Let $u(\l) = 1 - h(1-\l)$, and let $K$ be a shorthand for $c \log(n) \log \log(n)$. Then, since $h$ is increasing, so is $u$, and we have $u' \ge K u (1-u)$. Recall that Reed-Muller codes are $2$-transitive. Since $u$ is continuous in $\l$ we have, by the sharp threshold of $u$ at $R$, that for some $R^{\ast}$ with $|R^{\ast}- R| \le o_n(1)$ holds $u\(R^{\ast}\) = 1/2$, which also implies that $0 \le u(\l) \le 1/2$ for $0 \le \l \le R^{\ast}$. Hence in the interval $\left[0, R^{\ast}\right]$ we have that $u' \ge \frac12 K u$.

\noi Let $f(t) = u\(R^{\ast} - t\)$. Then $f(0) = \frac12$ and $f'(t) \le -\frac12 K f(t)$. Hence, by Gronwalls's inequality \cite{G}, we have that for $t > 0$ holds
\[
f(t) ~\le~ \frac12 \cdot \mathrm{exp}\left\{-\int_0^t \frac12 K ds\right\} ~=~ \frac12 \cdot \mathrm{exp}\left\{-\int_0^t \frac12 C \log(n) \log \log(n) ds\right\} ~=~ \frac12 n^{-\frac{Ct \log(n)}{2}}.
\]

\noi Since $u$ is increasing, this implies that $u(\l) \le o\(\frac{1}{\sqrt{n}}\)$ for $\l \le R^{\ast} - o_n(1)$. Recalling that $u = \mu'$ and that $|R^{\ast}- R| \le o_n(1)$, we have that for some $R^{\ast \ast}$ with $|R^{\ast \ast}- R| \le o_n(1)$ holds $\mu\(R^{\ast \ast}\) = o\(\sqrt{n}\)$.

\noi We can now conclude the proof. Let $C$ be a Reed-Muller code of rate $R$. Then $C^{\perp}$ is a Reed-Muller code of rate $1-R$, and hence by the preceding argument applied to $C^{\perp}$, we have that $\mu_{C^{\perp}}\(1-R^{\ast}\) = o\(\sqrt{n}\)$, where $|R^{\ast}- R| \le o_n(1)$. Let $\(a_0,...,a_n\)$ be the distance distribution of $C$. Recalling that he minimal distance of $C$ is $\Omega\(\sqrt{n}\)$, and applying Proposition~1.3 and Lemma~1.4 in \cite{S}, we have that for any $0 \le i \le n$ holds
\[
a_i ~\le~ 2^{o\(\sqrt{n}\)} \cdot \(\frac{1}{1-R^{\ast}}\)^{2 \ln 2 \cdot i} ~\le~ O\(\(\frac{1}{1-R^{\ast \ast}}\)^{2 \ln 2 \cdot i}\),
\]
where $|R^{\ast \ast}- R^{\ast}| \le o_n(1)$.

\noi We pass to the second claim of the proposition. We proceed as above, using the same notation. Let $C$ be a doubly transitive code of rate $0 < R < 1$ and let $0 < c \le 1$ be a given constant. By \cite{FK, Rossignol}, see also Section~3.1 in \cite{KKMPSU}, we have that
\[
\frac{dh}{dp} ~\ge~ k(p) \ln(n) \cdot h (1-h),
\]
where $k(p) \ge \frac{1-2p}{p(1-p) \ln\(\frac{1-p}{p}\)} - o_n(1)$. By Gronwall's inequality, this means that $u\(\frac R2\) \le n^{-t}$, for some absolute constant $t = t(R) > 0$.

\noi Similarly to \cite{KKMPSU}, we now use the fact that $h(p)$ is a measure w.r.t. the product measure $\mu_p$ of an increasing set $\Omega$ in $\{0,1\}^{n-1}$. Equivalently, $u(\l)$ is the measure w.r.t. the product measure $\mu_{\l}$ of an increasing set $\Omega^{\ast}$ in $\{0,1\}^{n-1}$, where $\Omega{^\ast} = \left\{x \in \{0,1\}^{n-1}, x \oplus 1 \in \Omega^c\right\}$ (here $\Omega^c$ is the complement of $\Omega$). We can now apply e.g., Lemma~2.7 in \cite{EKL}, to obtain that for any $b > 1$  holds
\[
u\(\(\frac{R}{2}\)^b\) ~=~ \mu_{\(\frac{R}{2}\)^b}\(\Omega^{\ast}\) ~\le~ \(\mu_{\frac{R}{2}}\(\Omega^{\ast}\)\)^b ~=~ u^b\(\frac{R}{2}\) ~\le~ n^{-bt}.
\]
This means that if $b > \frac{1-c}{t}$, for any $\l \le \(\frac{R}{2}\)^b$ holds $u(\l) \le o\(n^{c-1}\)$, which means that $\mu\(\(\frac{R}{2}\)^b\) \le o\(n^c\)$.

\noi We can now conclude the proof of the second claim of the proposition, similarly to the above, by applying the preceding argument to $C^{\perp}$, and by using the fact that the minimal distance of $C$ is $\Omega\(n^c\)$. We can choose $a = \(\frac{1-R}{2}\)^{\frac{1-c}{t}}$, where $t$ is given by $u_{C^{\perp}}\(\frac{1-R}{2}\) = n^{-t}$.

\noi We pass to the third claim of the proposition. Let $R_0$ be a sufficiently small constant, and let $R \le R_0$. We proceed as in the discussion above, using the same notation, but work directly with $C^{\perp}$. The function $u = u_{C^{\perp}}$ has a sharp threshold at $1-R$, implying in particular that $u\(1-R^{\ast}\) = \frac12$ for some $R^{\ast}$ with $|R^{\ast}- R| \le o_n(1)$. Moreover, on $\left[0,1-R^{\ast}\right]$ holds $u' \ge \frac12 \frac{1-2\l}{\l(1-\l) \ln\(\frac{1-\l}{\l}\)} \log(n) \cdot u$.

\noi Next we choose (with forethought) $R_1 = \(R^{\ast}\)^{e^{-8}}$, and consider the function $u$ on the interval $I = \left[1 - R_1, 1 - R^{\ast}\right]$. Choosing $R$ to be sufficiently small, we can ensure that $R_1$ is small enough to guarantee that for $\l \in I$ holds $u' \ge \frac14 \frac{1}{(1-\l) \ln\(\frac{1}{1-\l}\)} \ln(n) \cdot u$. Let $f(t) = u\(1 - R^{\ast} - t\)$. Then $f(0) = \frac12$ and for $t \in \left[0, R_1 - R^{\ast}\right]$ holds $f'(t) \le - \frac14 \frac{1}{\(R^{\ast}+t\) \ln\(\frac{1}{R^{\ast}+t}\)} \ln(n) \cdot f(t)$. Hence, by Gronwalls's inequality, we have
\[
f\(R_1 - R^{\ast}\) ~\le~ \frac12 \cdot \mathrm{exp}\left\{-\frac 14 \ln(n) \int_0^{R_1 - R^{\ast}} \frac{dt}{\(R^{\ast}+t\) \ln\(\frac{1}{R^{\ast}+t}\)} \right\} ~=~
\]
\[
\frac12 \cdot \mathrm{exp}\left\{-\frac 14 \ln(n) \cdot \(\ln \ln\(\frac{1}{R^{\ast}}\) - \ln \ln\(\frac{1}{R_1}\)\)\right\}  ~=~ \frac12 n^{-2}.
\]

\noi This means that $u\(R_1\) \le \frac12 n^{-2}$, and since $u$ is increasing, we have $u(\l) \le \frac12 n^{-2}$ for all $\l \le R_1$. We can now conclude the proof of the third claim of the proposition, similarly to the above.

\eprf

\subsection{Proof of Corollary~\ref{cor:RM_BSC}}

\noi Both claims of the corollary follow immediately from Proposition~\ref{pro:min-dist} and from the following technical lemma. (This lemma is probably well-known, so we relegate its proof to the Appendix.)

\lem
\label{lem:exp-comp-BSC}
Let $C$ be a linear code with weight distribution $\(a_0,...,a_n\)$ and assume that $a_i \le c^i$ for some constant $c > 1$. Assume also that the minimal distance of $C$ is at least $\omega\(log(n)\)$. Then for error $p$ such that $4p(1-p) < \frac{1}{c^2}$, $C$ corrects errors in $\mathrm{BSC}(p)$ with high probability.
\elem

\eprf

\section*{Acknowledgments}

\noi We are grateful to Or Ordentlich for many very helpful conversations and valuable remarks. We would also like to thank Nathan Keller for a very helpful discussion.

\section{Appendix}

\subsection{Proof of Lemma~\ref{lem:exp-comp-BSC}}

\noi The probability of error using $C$ in $BSC(p)$ is the same as the probability that a non-zero word in $C$ would be at least as close as zero to the noise vector (assuming w.l.o.g. that zero was transmitted). Let $z$ denote the noise vector. It is closer to $x \in C$ it than to zero iff it chooses at least $|x|/2$ coordinates in the support of $x$. By Sanov's theorem (\cite{Cover-Thomas}) the probability of this happening is (writing $Y$ for a binomial random variable $Y \sim \mathrm{Bin}(|x|,p)$, and $D\(\frac12||p\)$ for the Kullback-Leibler divergence between $\(\frac12, \frac12\)$ and $(p,1-p)$):
\[
\mathrm{Pr}\left\{Y \ge \frac{|x|}{2}\right\} ~\le~ \(|x|+1\)^2 \cdot 2^{-|x| D\(\frac12||p\)} ~=~ \(|x|+1\)^2 \cdot (4p(1-p))^{\frac{|x|}{2}}.
\]

\noi Let $P$ be the probability of error. Denoting by $d$ the minimal distance of $C$, and using the assumptions of the lemma, we have, via the union bound, that
\[
P ~\le~ O\(n^2\) \cdot \sum_{i=1}^n a_i (4p(1-p))^{\frac{i}{2}} ~\le~ O\(n^2\) \cdot \sum_{i=1}^n c^i (4p(1-p))^{\frac{i}{2}} ~\le
\]
\[
O\(n^2 \cdot \big(4c^2p(1-p)\big)^{\frac d2}\) ~\le~ o\(1\).
\]

\end{document}